# Ice-assisted soft-landing deposition for van der Waals integration

Xinyu Sun[123], Xiang Xu[13], BinBin Jin[4*], Yihan Lu[13], Jichuang Shen[13], Wei Kong[123*], Ding Zhao[123*], Min Qiu[123*]

*1*   *Key Laboratory of 3D Micro/Nano Fabrication and Characterization of Zhejiang Province, School of Engineering, Westlake University, 18 Shilongshan Road, Hangzhou 310024, Zhejiang Province, China.*

*2*   *Westlake Institute for Optoelectronics, Fuyang, Hangzhou 311421, China.*

*3*   *Institute of Advanced Technology, Westlake Institute for Advanced Study, 18 Shilongshan Road, Hangzhou 310024, Zhejiang Province, China.*

*4*   *School of Information and Electrical Engineering, Hangzhou City University, Hangzhou 310015, China.*

*Email:*
*jinbinbin@westlake.edu.cn;*
*kongwei@westlake.edu.cn;*
*zhaoding@westlake.edu.cn;*
*qiu_lab@westlake.edu.cn*

**Keywords:** van der Waals (vdW) integration, two-dimensional (2D) semiconductors, halide perovskites, ice buffer layer

**Abstract**

Van der Waals integration enables the creation of electronic and optoelectronic devices with unprecedented performance and novel functionalities beyond the existing material limitations. However, it is typically realized using a physical pick-up-and-place process to minimize interfacial damages and is hardly integrated into conventional lithography and metallization procedures. Here we demonstrate a simple in situ transfer strategy for van der Waals integration, in which a thin film of amorphous water ice acts as a buffer layer to shield against the bombardment of energetic clusters during metallization. After ice sublimation, the deposited metal film can be gently and in situ placed onto underlying substrates, to form an atomically clean and damage-free metal-semiconductor interface. This strategy allows ultra-clean and non-destructive fabrication of high-quality contacts on monolayer $MoS_2$, which is extremely beneficial to produce a high-performance 2D field-effect transistor with an ultra-high on/off ratio of $10^{10}$, mobility of 80 ($cm^2$ $V^{-1}s^{-1}$), and also with reduced Fermi level pinning effect. We also demonstrate the batch production of CVD-grown $MoS_2$ transistor arrays with uniform electrical characteristics. Such a gentle and ultra-clean fabrication approach has been further extended to materials with high reactivity, such as halide perovskites. Our method can be easily integrated with mature semiconductor manufacturing technology and may become a generic strategy for fabricating van der Waals contacted devices.

**Introduction**

Van der Waals (vdW) integration between semiconductors and metal electrodes with pristine interfaces is essential for unlocking novel device capabilities, advancing fundamental material studies, and achieving unprecedented performance.[1–5] However, integrating vdW contacts into conventional lithography and metallization procedures presents substantial challenges. In the case of two-dimensional (2D) semiconductors, which feature atomically thin bodies and flat surfaces with arbitrarily stacked layered structures, these challenges are particularly pronounced.[6,7] The metallization of 2D semiconductors is traditionally achieved through energetic metal deposition methods (e.g., e-beam/thermal evaporation and sputtering), which inevitably introduce defects, strain, and metal diffusion, resulting in interfacial chemical disorder and Fermi-level pinning.[8–14] The challenge of establishing reliable, high-quality vdW contacts with semiconductors emphasizes the importance of addressing these obstacles to realize the full potential of next-generation semiconductor technologies.[15]

To make high-quality vdW contacts of 2D semiconductors upon metallization, transfer printing technology has been proposed and employed to demonstrate state-of-the-art field-effect transistors (FETs) crafted from transition metal dichalcogenide (TMDC) films.[12,16–19] In particular, $MoS_2$ transistors with transferred electrodes exhibit excellent on/off ratios of $10^6$~$10^9$,[12,20–24] highlighting the considerable potential of 2D semiconductors. High-quality metal-semiconductor contact relies on the pre-deposition, physically peeled off with a stamp holder and transferred onto 2D semiconductors.[11,15,20] Accordingly, the intrinsic physicochemical properties of 2D semiconductors could be well retained, resulting in ideal vdW contacts. However, this method is available only for a few low-adhesion metals, while failure in transferring metal with strong adhesion. Alternatively, a graphene-assisted metal transfer printing technique was proposed to transfer arbitrary metal electrodes.[10] Attributed to the weak vdW surface of graphene, metal electrodes could be easily delaminated and printed onto 2D semiconductors. Nonetheless, certain challenges persist during the pick-up-and-place transfer processes, such as the occurrence of undesired structural damage, crumbling, wrinkles, and the formation of interfacial bubbles.[1,26–28] Additionally, the procedure for releasing the electrode windows, which necessitates the removal of the stamping holder, conventionally entails the utilization of chemical solutions and/or heating treatments.[29,30] This process results in the inescapable presence of chemical residues, which possess the potential to degrade and modify device performance.[27] Moreover, it imposes limitations on the applicability of the technique to specific material systems.

Most recently, a wafer-scale vdW integration strategy has been achieved by utilizing thermally decomposable polymer (poly (propylene carbonate), PPC) as the buffer layer between metal and 2D semiconductors.[11] The polymer layer was dry-decomposed at 250 °C into gases, leading to the metal free settling on the 2D semiconductors with the sharp interface. These methods demonstrate the ability to transfer arbitrary metal electrodes onto 2D semiconductors to realize vdW contacts, however, the specific operational process poses a critical limitation for their application extended to low thermal resistance material, such as widely used organic-inorganic halide perovskites (OIHPs) and organic semiconductors which are pivotal for advancing energy conversion as well as photonic detection and sensing.[31–34] The high reactivity commonly associated with OIHPs and organic semiconductors poses a significant challenge to the development of reliable contacts. Moreover, the carbonyl group in PPC has strong adsorption capacity to 2D materials, which indicates the potential presence of organic residues at the interface. Therefore, it is essential to develop a general vdW integration technique not only limited to 2D semiconductors but could be well-extended to fragile and reactive materials with poor thermal stability. [35,36]

Here, we introduce a simple and universally applicable method for vdW integration. In this approach, an amorphous ice layer serves as a buffer, preventing the bombardment of energetic metal clusters and allowing the deposited film to softly land onto semiconductors during ice sublimation, leading to a vdW contact. Different from the typical transfer "pick-up-and-place" processes from wafer to wafer, we exploit a much simpler and high-efficiency "in situ transfer" process that eliminates contamination and damage and is also suitable for large-scale production. As verified by detailed characterization, the resulting vdW interface is damage-free and atomically clean. Using this technique, our MoS$_2$ FET devices reveal ultrahigh on-off ratio of $10^{10}$, mobility of 80 cm$^2$V$^{-1}$s$^{-1}$ and reduced Fermi level pinning effect. As a proof of concept, we fabricate an array of MoS$_2$ FETs on a large scale, which exhibit high uniformity. Furthermore, taking advantage of the chemical residue-free and solvent-free process, this strategy has been successfully extended to fabricate OIPH devices with vdW contact, while also demonstrating enhanced photodetector performance. Such an ultra-clean fabrication approach could be the ideal platform for vdW integration for advanced optoelectronics applications.

**Ice-assisted vdW integration**

The ice-assisted vdW contact takes advantage of amorphous ice serving as a buffer layer, hence, energetic metal clusters bombard directly onto the ice by thermal evaporation. During the dissipation of the buffer layer, the deposited film is physically in situ laminated onto pristine 2D semiconductors in vacuum. This process effectively sidesteps the chemical involvement (e.g., lift-off step using polar solvent in standard lithography process or interfacial residues brings by transfer stamps) and removes restrictions on the choice of deposited materials or substrates. The schematic fabrication process flow is illustrated in **Fig. 1a**. MoS$_2$ flakes are mechanically exfoliated on the silicon substrate covered with SiO$_2$. The sample, together with a shadow mask, is positioned on a home-made cryogenic stage and cooled to below 130 K (**Fig. 1a (i)**). Subsequently, water vapor is sprayed onto the sample to create a uniform film of amorphous ice with a thickness of around 90 nm (**Fig. 1a (ii), Supplementary Fig. S1 and S2**), followed by deposition of an Au layer (**Fig. 1a (iii)**). Finally, the sample is heated to room temperature in vacuum. The sublimation process of the ice layer, initially starting at the electrode edges and gradually progressing until complete sublimation, is schematically illustrated in **Fig. 1b-c**. At this stage, the ice is completely sublimated, leaving the top metal film naturally laminated onto the MoS$_2$ substrate (**Supplementary Discussions 1**). In a cross-sectional high-resolution transmission electron microscopy (HRTEM) image, we can clearly see an atomically sharp, clean and damage-free interface between the Au lattice and MoS$_2$ lattice (**Fig. 1d**), whereas the directly deposited Au-MoS$_2$ interface is highly disordered, with considerable defects, metal diffusion and formation of a winding interfacial layer (**Fig. 1e-g**). Furthermore, we show the cross-sectional scanning transmission electron microscopy image (STEM) of the vdW interface in a larger scale (**Fig. 1h**) and energy dispersive spectrometer (EDS) elemental analyses (**Fig. 1i**). The TEM and element analysis studies clearly prove that our vdW integration approach offers an ultra-clean and damage-free integration process. Moreover, ice-assisted vdW integration has demonstrated excellent compatibility with high-adhesion metals, such as Cr (**Supplementary Discussions 2**).

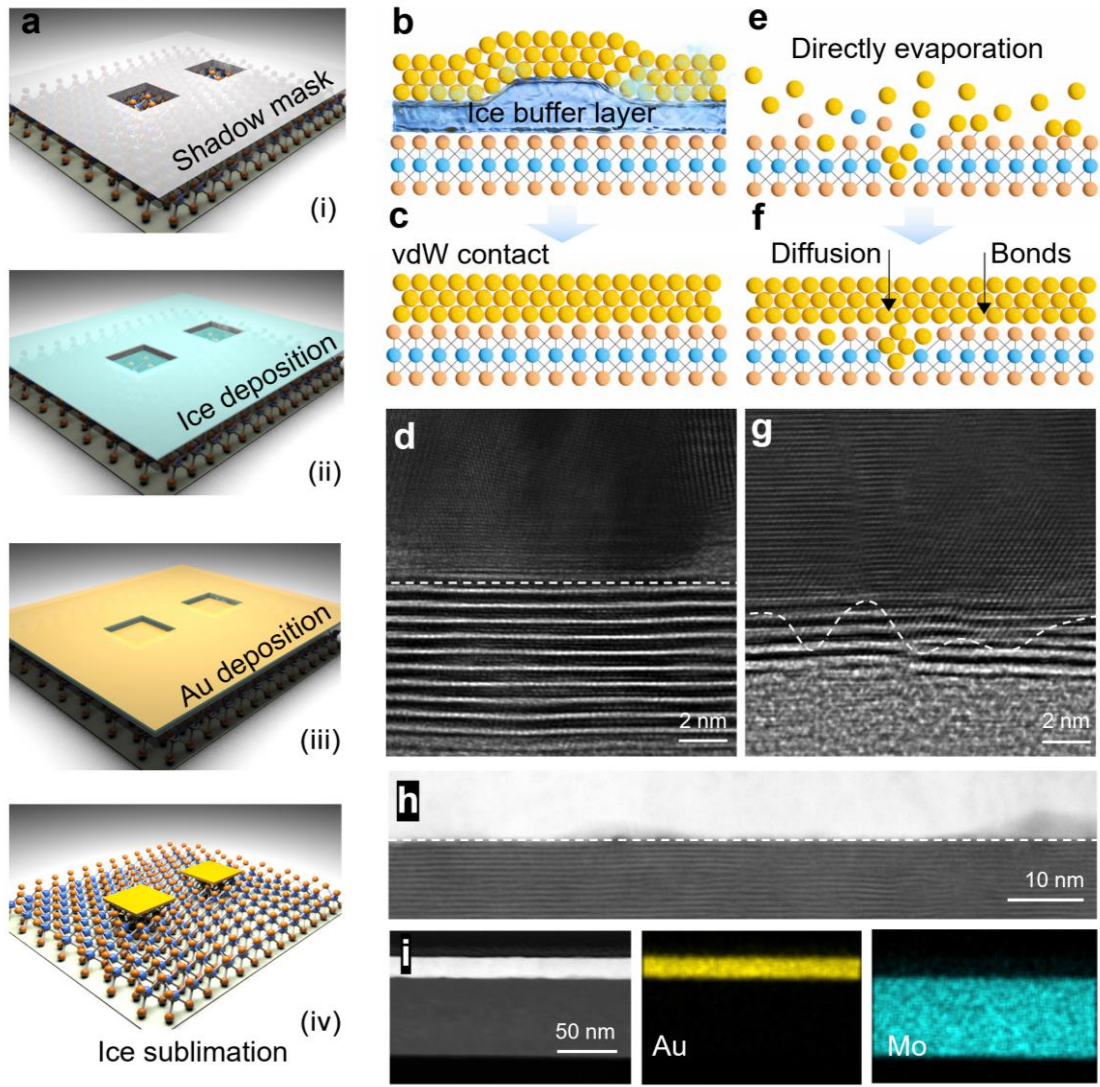

**Fig. 1 Fabrication process and comparison of vdW contact with conventional evaporated contact. (a)** Schematic steps include shadow mask placement, ice deposition, metal deposition, and ice sublimation. **(b–d)** Cross-sectional schematics and high-resolution transmission electron microscopy (HRTEM) image of the ice-assisted Au electrode on top of $MoS_2$, emphasizing atomically sharp and clean metal-semiconductor interfaces, highlighted by the straight dotted line. **(e-g)** Cross-sectional schematics and HRTEM image of conventional thermal deposition Au electrodes on top of $MoS_2$, interfacial layer highlighted by the curved dotted lines. The bombardment of the $MoS_2$ surface by high-energy Au atoms and clusters creates considerable damage to the $MoS_2$ surface, resulting in damage, defects, interface diffusion, and atomic disorder. **(h)** Scanning transmission electron microscopy (STEM) image of the cross-section of the Au-$MoS_2$ junctions vdW interface in a larger scale, where the contact shows a straight interface. **(i)** The STEM image and element analysis of Au-$MoS_2$ junctions vdW interface.

To further demonstrate the weak vdW interaction at the interface, we mechanically separated the metal electrodes from $MoS_2$ after the device fabrication and electrical measurement, which could provide robust indicators for ideal interfaces.[8] The underlying $MoS_2$ maintained its original shape without any discernible damage (**Fig. 2a and b**). Conversely, for the evaporated contact (EVC) electrodes upon mechanically peeling off, the underlying $MoS_2$ underwent simultaneous destruction (**Fig. 2c and d**), signifying the tendency of deposited metal electrodes to typically form robust chemical bonds with the underlying $MoS_2$, such as Au–S bonds.[37] The reversible physical integration and isolation of the Au–$MoS_2$ junction benefit from our ice-assisted process also prove that the two materials maintain their isolated states without direct chemical bonding.

In the as-fabricated Au–$MoS_2$ junction, thanks to the weak vdW interaction, allowing for the mechanical separation of integrated Au electrodes. This separation facilitates the inversion of the Au film's bottom surface for atomic force microscopy (AFM) measurements (**Fig. 2e**), which could be a complement to characterize the interface properties of the larger contact region in two dimensions. **Figure 2f** shows that the peeled Au's bottom surface is atomically flat, with a root-mean-square (RMS) roughness of 0.38 nm, mirroring the flat surface of the ice film. Simultaneously, the decoupled $MoS_2$ exhibits a flat surface with a small RMS roughness of 0.25 nm (**Fig. 2g**), consistent with the surface quality previously measured by AFM.[38] The uniformity and flatness observed in both the Au bottom surface and $MoS_2$ top surface, following the separation of the fabricated interface, provide compelling evidence of the optimized vdW interface across the entire contact region. In addition to the AFM analysis of the ice-assisted vdW interface, photoluminescence (PL) and X-ray photoelectron spectroscopy (XPS) were conducted to investigate the impact of the ice-assisted process on the material and the devices (**Supplementary Discussions 3 and Fig. S3**). The results indicate that the material retained its original properties after ice deposition and sublimation.

We emphasize the crucial role of water ice in our vdW contact approach. Capitalizing on its unique properties, it has contributed to an extremely simple and eco-friendly process. The method eliminates contamination from solvents and chemical components, addressing the issue of chemical residues at its source. This makes the approach particularly suitable for creating vdW contacts on halide perovskite thin films, which are generally soluble in various solvents and incompatible with traditional lithography processes. The "soft-lattice" structure is also susceptible to degradation during metal deposition process.[39] Here, this ice-assisted strategy was adopted to construct vdW contacts on OIHP materials. To implement the ice-assisted vdW contact strategy while mitigating the impact of deposition conditions,[40] we executed a series of processes for the evaporation of Au electrodes onto polycrystalline $CH_3NH_3PbI_3$ ($MAPbI_3$) films, including ice-assisted process, thermal evaporation under cryogenic deposited process and normal thermal evaporation process. We then attempted to peel off the electrodes through blue membrane to expose the underlying perovskite surfaces. The vdW contact device was easily detached due to the weak vdW force between Au electrodes and $MAPbI_3$ (**Fig. 2h**). In contrast, the peeling-off yield of the cryogenic evaporated contacts (Supplementary **Fig. S5**) was significantly lower, and no pair of electrodes were detached in the case of normal evaporation process (**Fig. 2i**). This discrepancy suggests a much stronger chemical bonding with underlying perovskite. Notably, in situations where the cryogenically deposited contacts were detached, the residual Au was still detected by XPS, while for vdW contacts no Au signal was detected (Supplementary **Fig. S6**). The PL mapping results also revealed that the emission intensities of $MAPbI_3$ initially covered by cryogenic deposition process were much lower than that of those covered by the vdW deposition process, indicating the perovskite lattice was seriously damaged by high-energy metal bombardment. Furthermore, we extended this vdW contact strategy to deposit Au electrodes on single-crystal $MAPbBr_3$ nanoplates

(Supplementary **Fig. S7**). For ice-assisted vdW contacted sample, Au electrodes are ready to be peeled off with negligible PL emission variation of underlying MAPbBr$_3$ nanoplate, while for conventional evaporated contacted sample, electrodes cannot be peeled off. To demonstrate the performance enhancement of vdW contacts for perovskites, we used the ice-assisted process to create a vdW-integrated electrode-perovskite interface in a MAPbBr$_3$ photodetector (**Supplementary Discussions 4**). The vdW contact MAPbBr$_3$ photodetector shows lower dark current ($I_{dark}$) and higher photocurrent ($I_{photo}$) under varying light intensities, demonstrating the improved photodetector performance of the vdW-integrated interface. In evaporated contact devices, higher trap density at the Au-MAPbBr$_3$ interface increases trap-assisted nonradiative recombination, while ion migration reduces charge collection, enhancing bimolecular recombination loss.

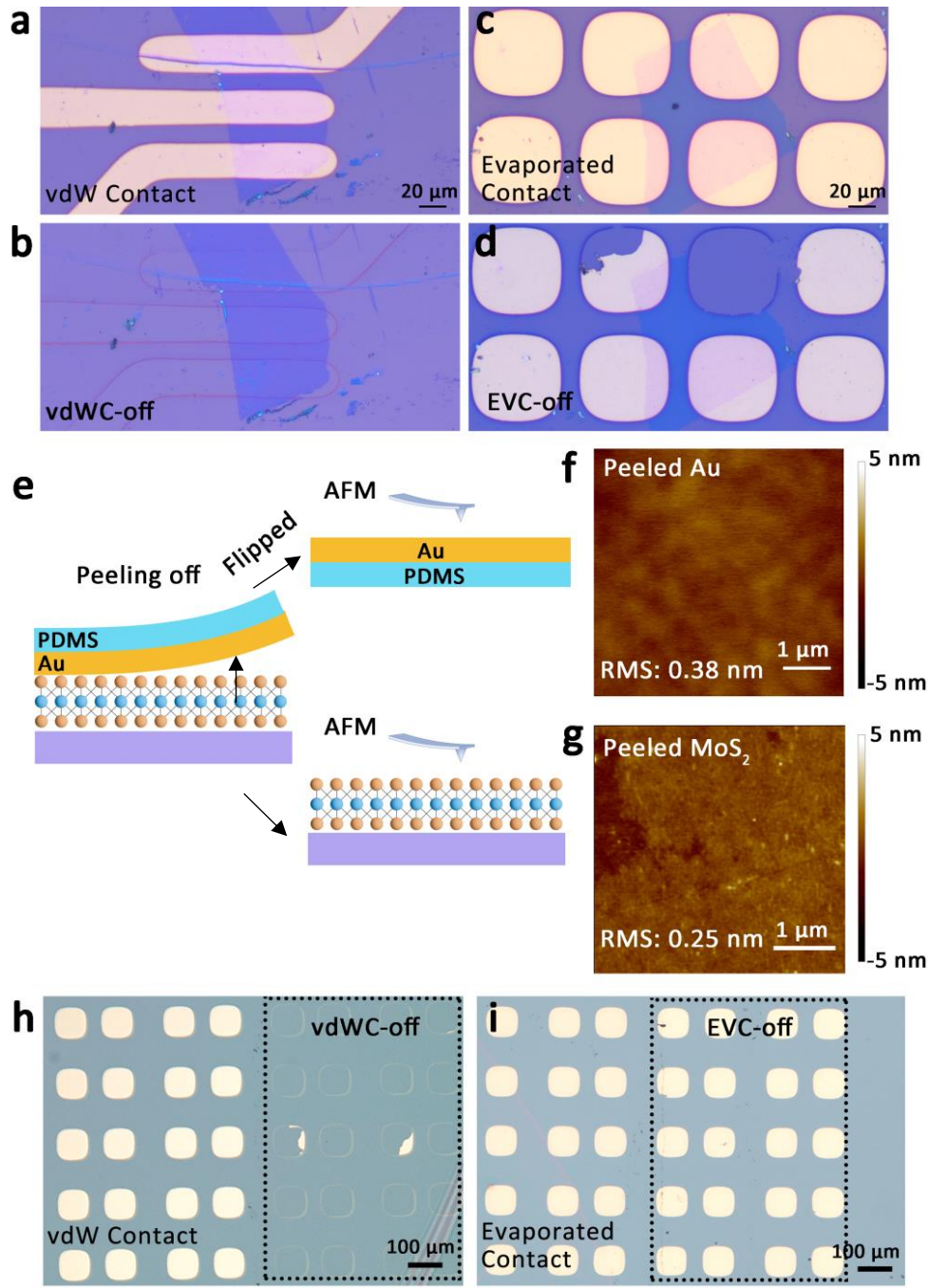

**Fig. 2 Characterizations of vdW interface. (a-b)** Optical images of the MoS$_2$ device with ice-assisted vdW contact electrodes and mechanically released electrodes, confirming the preservation of the MoS$_2$ layer's original shape. **(c-d)** Optical images of a MoS$_2$ device with directly deposited electrodes, revealing destruction during electrode removal, suggesting strong chemical bonding in deposited junctions between the MoS$_2$ and Au. **(e-g)** Schematic depiction of the AFM sample preparation process for assessing interface quality on a large scale. The Au/MoS$_2$ interface is decoupled by peeling off the Au electrodes. AFM characterization of the flipped Au bottom surface **(f)** shows an RMS of 0.38 nm, and AFM characterization of the MoS$_2$ top surface after Au electrode removal **(g)** exhibits an RMS of 0.25 nm. **(h-i)** Optical images of fabricated electrodes on MAPbI$_3$ thin films of vdW contact (vdWC) and evaporated contact (EVC) before and after peeling off (refer to the area of vdW-off and EVC-off).

**Electrical properties of vdW-integrated MoS$_2$ transistors**

To demonstrate the improved electrical performance provided by our ice-assisted vdW contact, we conducted electrical property measurements on the exfoliated monolayer MoS$_2$ transistors with vdW contact Au electrodes (Device1). A highly doped Si (p++) substrate served as the back gate, and a 285 nm thick SiO$_2$ was employed as the gate dielectric (**Method section**). The source and drain electrodes were composed of vdW contact Au electrodes, and the channel length was consistently set at 15 μm, as defined by the shadow mask. For comparison, MoS$_2$ transistors with thermal evaporation under cryogenic (Device2) and conventional thermal evaporation (Device3) were fabricated, respectively. In these cases, an identical channel material and the same channel length were utilized for a fair and accurate comparison. Notably, Device1 and Device2 were fabricated on the same monolayer of MoS$_2$ of distinct regions, as illustrated by the optical images in the inset of **Fig. 3**.

The output curves of the three devices measured at room temperature are exhibited in **Fig. 3a-c**, and the transfer curves are exhibited in **Fig. 3d-f**. All devices show a linear drain-source current versus drain-source voltage ($I_d$-$V_d$), while the major difference is their on-state current, compared with the evaporated contact devices, the vdW contacted sample (Device1) possesses highest on-state current ($I_{on}$). The extracted $I_{on}$ of vdW contact under $V_g$ of 40V is 6.64 μA/μm ($V_d$ of 3V) and 2.28 μA/μm ($V_d$ of 1V), 4.3 times higher compared to Device2 with 1.53 μA/μm ($V_d$ of 3V), and 76 times higher compared to Device3 with 0.03 μA/μm ($V_d$ of 1V). It is noteworthy that the on-state currents exhibit a close correlation with barrier heights. In the case of a metal contact featuring a higher barrier height, the transport of electrons from the metal contact to the conduction band of MoS$_2$ is impeded, leading to a reduction in the saturated on-state current of the transistor.[11]

The transfer curves of the three devices all exhibit n-type transfer characteristics. Despite the theoretical prediction based on defect-free MoS$_2$ transistors suggesting p-type doping, it is widely accepted that abundant sulfur vacancies are prevalent as native defects in MoS$_2$ bulk crystals under thermodynamic equilibrium, making p-type doping challenging.[10,41,42] We also report the forward and reverse transfer curves for three devices, plotted both on logarithmic and linear scales. Device1 exhibits slighter hysteresis transfer curves across the whole range of $V_g$ compared with Device2 and Device3. For vdW contacted Device1, the extracted carrier mobility is 66.8 cm$^2$V$^{-1}$s$^{-1}$, which is much higher than evaporated contact Device2 with extracted mobility of 9.28 cm$^2$V$^{-1}$s$^{-1}$ and Device3 with 3.4 cm$^2$V$^{-1}$s$^{-1}$. The normalized on-state current values $I_{on}$L/W and subthreshold swings are 67 μA, and 500 mV per decade respectively. Additionally, Device1 showed an ultrahigh on/off ratio of 6.3× 10$^9$. In contrast, the on/off ratio of the evaporated contact device (Device3) was significantly reduced to 10$^7$, making a three-orders-of-magnitude reduction. Since the same channel material is used, the observed much better electrical properties could be largely attributed to the optimized contact through our vdW integration method, minimizing the defects and trapped charges at the surface of the MoS$_2$ or in the MoS$_2$ itself, and the reduced contact resistance.

To demonstrate the robust and tunable device characteristics, we conducted temperature dependent measurements of vdW contact devices with different metals. Temperature-dependent transfer characteristics obtained at $V_d$=2V of vdW contact FETs with Au electrode (Device1) and Ag electrode (Device4) are depicted in **Fig. 3g-h**. The vdW contact FETs both exhibit an ultrahigh on/off ratio of ~10$^{10}$ at 4.9 K. In the examination of the temperature-dependent transfer characteristics of metal Au (with a work function of 5.1 eV) and Ag (with a work function of 4.3 eV), it was observed that the temperature-dependent current variation (indicated by dashed black arrows) in the Au vdW contact FET (Device1) exceeded that in the Ag vdW contact FET (Device4), which means a large Schottky barrier height (SBH) obtained Device1. The Ag vdW contact FET (Device4) also exhibits even higher mobility of 80 $cm^2V^{-1}s^{-1}$ and 274 $cm^2V^{-1}s^{-1}$ at 298K and 4.8K, respectively. Additionally, **Fig. S12a** shows the temperature-dependent output characteristics of the Device4. It unveils linear $I_d$–$V_d$ output curves that persist even at 4.8 K, suggesting the formation of good ohmic contact with $MoS_2$ and a low SBH.[44] In contrast, the temperature-dependent output curves of the vdW contact FET with Au (Device1) show nonlinear characteristics **Fig. S12b**.

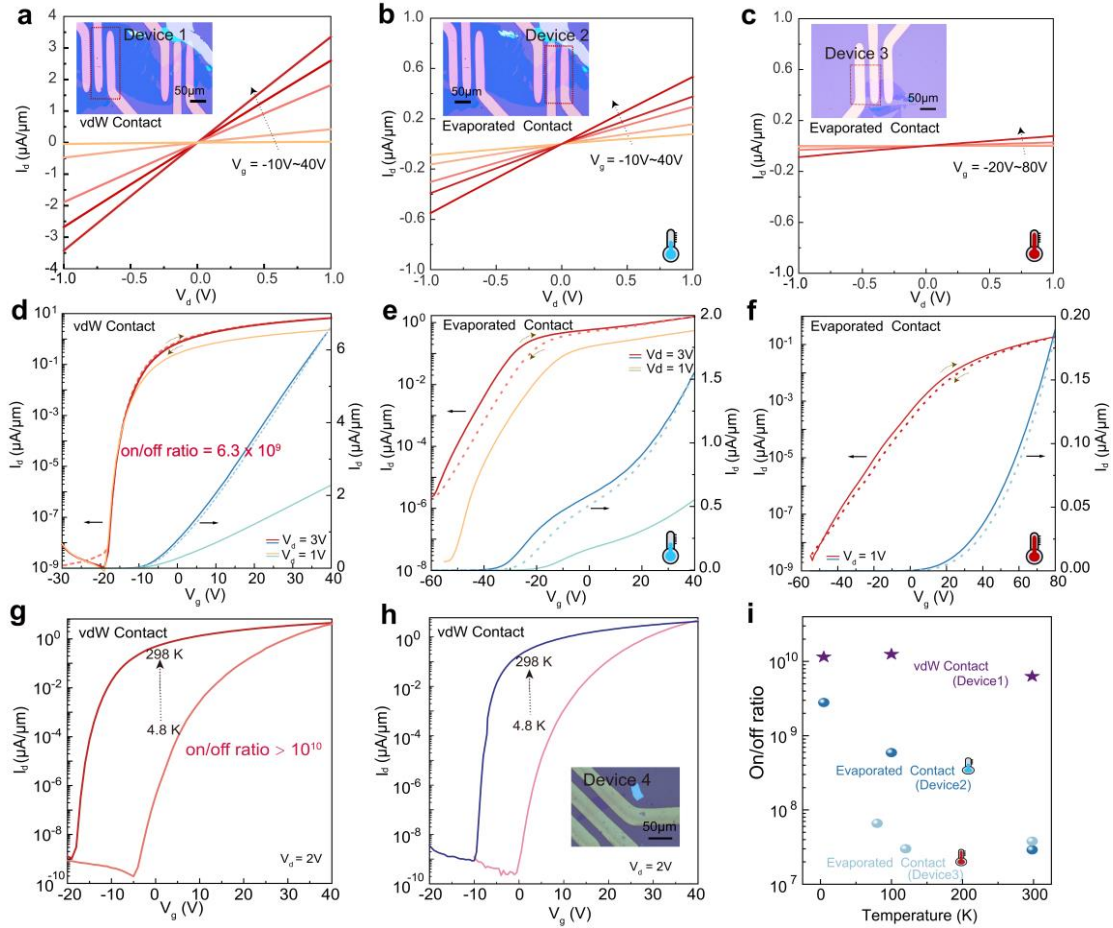

**Fig. 3 Electrical Performance of MoS₂ FET. (a)** Room-temperature output curves ($I_d$–$V_d$) with different $V_g$ values of the $MoS_2$ FETs using vdW-contact Au electrodes (Device1), **(b)** evaporated-contact Au electrodes fabricated under cryogenic condition (Device2), and **(c)** evaporated-contact Au electrodes under normal condition (Device3). Insets: optical images corresponding to the $MoS_2$ FET used in our measurements, fabricated using different methods. **(d-f)** Room-temperature transfer curves ($I_d$–$V_g$) of Device1 and Device2 measured at $V_d$ =1V and 3V, and Device3 measured at $V_d$ =1V. The forward (red line) and reverse (red dashed line) transfer curves are presented on logarithmic and linear scales. **(g-h)** Temperature-dependent transfer curves of the FETs with vdW contact Au electrodes (Device 1) and vdW contact Ag electrodes (Device 4), are measured at temperatures of 298 K and 4.8 K. Insets: optical image of the vdW contact Ag FET. **(i)** Comparisons of the on/off ratio versus temperature for the three devices.

To quantitatively evaluate the SBH, we performed measurements on ice-assisted vdW contact FETs using four metals with varying work functions (Ag, Cr, Cu, Au) by analyzing their temperature-dependent transport characteristics over a specified temperature range (**Supplementary Discussions 5**). The flat-band SBH of the Ag vdW contact device is determined to be around 65 meV, which is significantly lower than most reported Schottky barrier height values.[9,45,46] For comparison, the SBH values of Cr, Cu, and Au vdW contact FETs exhibit an increasing trend, further confirming that the SBH of our vdW-integrated metal contacts can be effectively tuned by selecting metals with different work functions.[9] The tunable device characteristics dictated by the metal work functions confirm our minimized damage and weak interaction in the interface.[8] In addition to evaluating the SBH, the contact resistance ($R_C$) of different metal contacts was measured using the transfer length method to further confirm the optimized contact in our ice-assisted vdW interface (**Supplementary Discussion 6**). The $R_C$ value for Ag contacts is determined to be 0.745 kΩ μm, consistent with previously reported high-quality contacts.[10,26,47] This low $R_C$ is attributed to a residue-free metal/$MoS_2$ interface that effectively mitigates Fermi pinning. When high work function metals, such as Cu and Au, are integrated via vdW methods, significantly higher $R_C$ values are observed. The discrepancy of SBH and contact resistance could be well-explained by the thermionic emission of electrons between the metal and $MoS_2$, where a lower work function metal forms a lower barrier.[48,49] The transfer characteristics of the CVD-grown $MoS_2$ transistors are presented in **Fig. S11e**, demonstrating that metal-dependent electrical behavior arises from $R_C$ variations among different metals.

Furthermore, we summarize the on/off ratios at three temperatures (4.9 K, 100 K, and 298 K) for all contact devices in **Fig. 3i**. Notably, the vdW contact FET (Device1) consistently maintains a high on/off ratio compared to the evaporated contact FETs (Device2 & 3), reaching an ultrahigh on/off ratio above $10^{10}$ at temperatures below 100 K. Moreover, the evaporated contact FET (Device2) also achieves an on/off ratio of $10^9$ at 4.9 K, significantly outperforming Device3. This underscores the importance of cryogenic conditions in the evaporation process for enhancing electrical properties. We remark that, as reported in the detailed comparison with the literature (**Supplementary Discussions 8, Table. S1**), our ice-assisted vdW-contact device features a state-of-the-art ultrahigh on/off ratios combined with on-current. Meanwhile, we emphasize that the achieved on/off ratio and mobility can indeed be further improved by utilizing PMMA or other high-k dielectric layers to mitigate the abundant trap states at the $SiO_2/MoS_2$ interface.[50] Reducing these trap states is also crucial for preserving the nearly intrinsic electronic properties of $MoS_2$.[29]

**Uniformity and performance of vdW-integrated CVD-grown $MoS_2$ FETs**

We next evaluated the uniformity and performance of ice-assisted vdW-integrated FETs using CVD-grown $MoS_2$,[51,52] transferred via wet transfer from sapphire to a 20 nm $Al_2O_3$/p++ substrate. To systematically assess the scalability and uniformity of the vdW integration process, we compared the homogeneity and electrical properties of $MoS_2$ transistors with ice-assisted vdW contacts to those of devices with evaporated contacts. The $MoS_2$ channels were patterned into ribbon-like areas, isolated by lithography and dry etching, with all devices sharing a consistent geometry featuring a 15 μm × 20 μm channel. **Figure 4a** presents photographs and optical images of the ice-assisted vdW contact FETs fabricated on CVD-grown $MoS_2$ with Ag contacts. Transfer characteristics from 75 randomly selected ice-assisted vdW-contacted $MoS_2$ FETs exhibit uniform n-type behavior with minimal device-to-device variability (**Fig. 4b**). In contrast, transfer characteristics from 75 randomly MoS$_2$ devices with evaporated contacts display substantial variability across transfer characteristics. The statistical distributions of on-state current (I$_{on}$) and off-state current (I$_{off}$) are shown in **Fig. 4c** and **Fig. 4d**, respectively. Devices with vdW contacts demonstrate a narrow distribution in I$_{on}$, averaging 60.6 μA, while devices with evaporated contacts exhibit a broader and generally lower I$_{on}$ distribution, averaging 21. 3 μA. This disparity stems from uncontrolled damage and interfacial defects introduced during the metal deposition process, which significantly affect the on-state performance. Furthermore, vdW contacted devices exhibit a substantially lower I$_{off}$ (<10$^{-12}$ A) relative to evaporated-contact devices (range from 10$^{-11}$A to 10$^{-8}$ A), representing an order-of-magnitude improvement in off-state performance and underscoring the suitability of vdW contacts for low-power applications. Statistical analysis of the on/off current ratio further reveals an average value of 9.6×10$^8$ for vdW devices, with a peak on/off ratio reaching 5.98×10$^9$ (**Fig. 4e**), demonstrating a competitive performance among reported CVD-grown monolayer MoS$_2$ transistors. By contrast, evaporated contact devices exhibit a notably lower on/off ratio, averaging 3×10$^8$, attributable to increased interfacial trapping states introduced during vacuum deposition and metal contact formation. Finally, field-effect carrier mobility and threshold voltage (V$_{th}$) data of vdW-contacted devices show average values of 47 cm² V$^{-1}$ s$^{-1}$ and 3.5 V, respectively. In contrast, evaporated contact devices exhibit significantly lower mobility (**Fig. 4f**) and higher threshold voltages (**Fig. 4g**). This comparison further underscores the advantages of ice-assisted vdW integration for achieving high-performance MoS$_2$ transistors with enhanced uniformity and scalability.

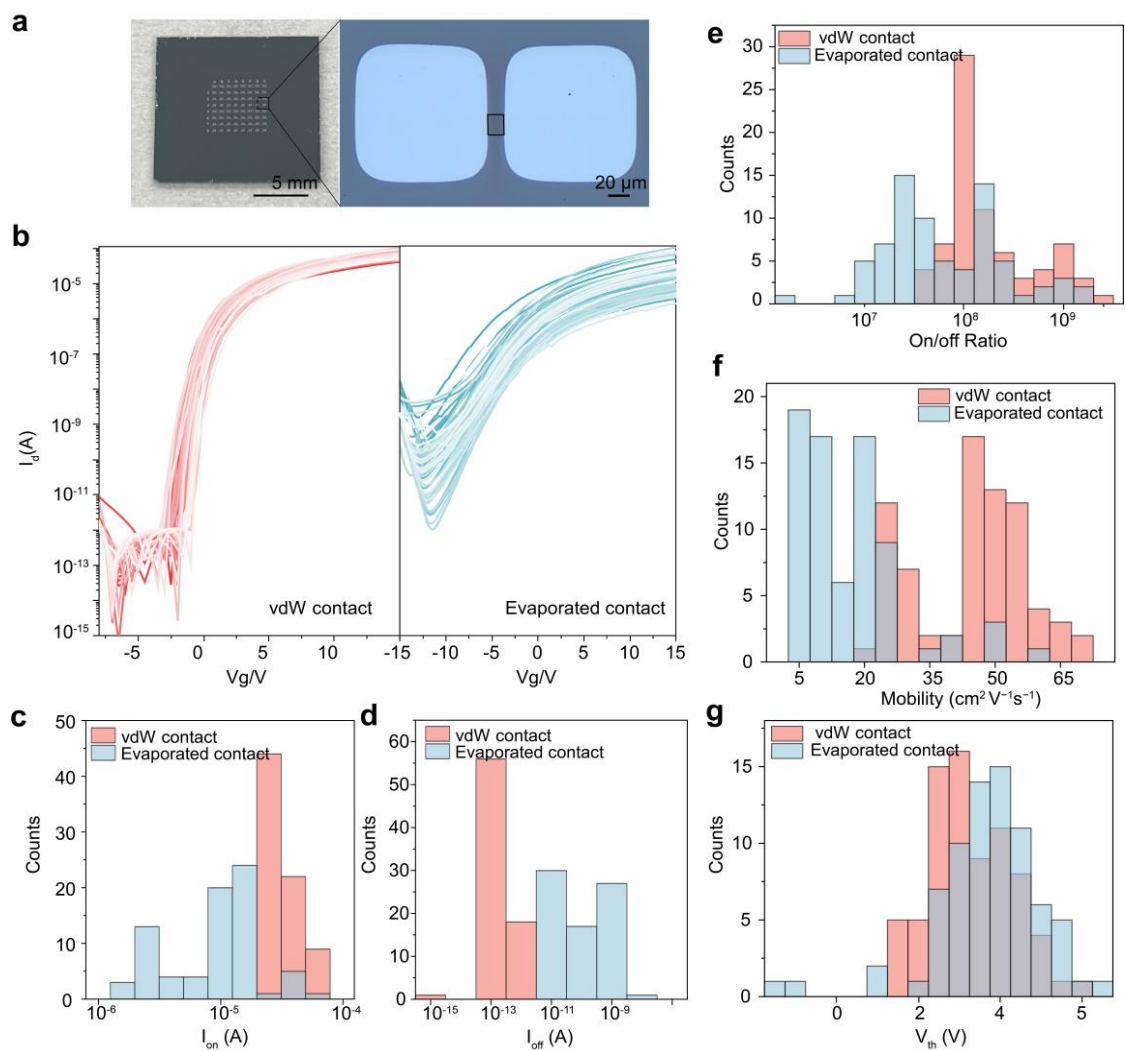

**Fig. 4 Electronic properties of batch-fabricated vdW-integrated monolayer CVD-MoS$_2$ FETs with Ag contacts. (a)** Optical image of batch-fabricated ice-assisted vdW contact MoS$_2$ FETs on 1.5 cm×1.5 cm Al$_2$O$_3$/p++ substrate and the optical microscope image of the single MoS$_2$ FET. The channel width and length are 20 μm and 15 μm**. (b)** Transfer characteristics (V$_d$ = 2 V) of 75 randomly selected MoS$_2$ FETs with vdW Ag contacts and 75 randomly selected devices with evaporated Ag contacts. **(c)-(g)** The comparison of statistical distribution of I$_{on}$ (c), I$_{off}$ (d), the on/off ratio (e), the mobility (f) and V$_{th}$ (g) at room temperature.

**Conclusion**

In summary, we present a simple, universal, and efficient approach for constructing vdW junctions through soft-landing metal electrodes on the material surface using water ice as a buffer layer. The water ice layer serving as a protective barrier against the harsh effects of hot metal bombardment. During a straightforward ice sublimation process, the metal is gently laminated onto the material surface, resulting in an atomically clean interface. This ice-assisted process avoids chemical material usage, preventing chemical residues at the source. In contrast to previous vdW integration methods, our strategy does not rely on metal adhesion and can be readily applied to various semiconductor systems, even organic-inorganic halide perovskites. Meanwhile, our method also mitigates damage, crumbling, wrinkles, and bubbles occurring in conventional metal transfer processes, particularly relevant in wafer-to-wafer processes. Furthermore, we show that our vdW integration process significantly enhances the performance of monolayer MoS$_2$ transistors. Our process enables integration of high-quality contacts on MoS$_2$, ultimately facilitating the creation of high-performance transistors with ultrahigh on/off ratios, higher normalized $I_{on}$, and reduced fermi pining effect compared to those achieved by conventional metallization. We also demonstrate the batch production of CVD-grown MoS$_2$ transistor arrays with uniform electrical characteristics. This strategy has the potential to bring a paradigm shift in addressing fundamental challenges in vdW integration and also serves as a promising platform for researching new materials and high-performance devices.

## Methods

**Fabrication processes for electrodes vdW integration and electrodes mechanically peeling off.**

The detailed descriptions of the materials used in various parts of this study are provided in **Supplementary Discussions 7**. The exfoliated monolayer MoS$_2$ sample was prepared by mechanical exfoliation onto a SiO$_2$ (285 nm)/p++ wafer. For devices using CVD-grown MoS$_2$, monolayer MoS$_2$ was transferred via wet transfer from a sapphire substrate to an Al$_2$O$_3$ (20 nm)/p++ substrate. Subsequently, the sample with a stencil mask is positioned on the cryogenic sample holder and cooled down to the required temperature (<130 K) in the scanning electron microscope (SEM) chamber (Supplementary **Fig. S16**). The accuracy of this temperature is confirmed by directly measuring the temperature of the holder and observing the state of deposited ice on the sample. Water vapor is then injected into the SEM and sprayed onto the sample to create a uniform film of amorphous ice with a thickness of 90 nm. The thickness of the ice is controlled by a water vapor injector (Supplementary **Fig. S2**). Following this, the cryogenic sample holder is transferred to the metal deposition chamber within 1 minute. In this chamber, 40 nm-thick electrodes are evaporated through a thermal evaporation process under vacuum conditions (pressure ~2×10$^{-4}$ Pa). After the metal deposition process, the sample is heated to room temperature under vacuum. The ice is heated up in a vacuum environment with a pressure of less than 1×10$^{-4}$ Pa, significantly lower than the triple point pressure of water. Consequently, the ice undergoes direct sublimation without transitioning through the liquid phase and is evacuated by the pump system.[53] At this stage, the ice is completely sublimated, leaving the top metal film naturally laminated onto the MoS$_2$ substrate.

Device1 and Device2 were fabricated sequentially. For Device1, ice layer was deposited after the initial cooling, followed by thermal evaporation. The adjacent MoS$_2$ for Device2 was masked to avoid exposure to both the ice and metal deposition during this stage, thereby omitting the ice layer deposition. During the fabrication of Device2, both devices underwent a second cooling cycle. At this stage, Device1 was masked, and metal deposition was performed on Device2. For the fabrication of Device2, the monolayer MoS$_2$ sample, along with a shadow mask, is directly placed on the sample holder in the metal deposition chamber and cooled to below 130 K. For the fabrication of the Device3, the sample is positioned directly on the sample holder in the room-temperature metal deposition chamber. The evaporation rate remains consistent across all fabrication methods.

To peel off the metal films on MoS$_2$, a 100 μm-thick PDMS layer was applied atop the metal. To peel off the metal films on perovskite, blue membrane was used to expose the underlying perovskite surfaces.

**Device characterizations.**

The electrical measurements were conducted using a Keithley 4200A-SCS source measurement unit (SMU) in a Lakeshore CRX cryogenic probe station. Measurements were taken at room temperature in a vacuum environment (pressure ~$5 \times 10^{-5}$ Torr) and at low temperature in a vacuum environment (pressure ~$5 \times 10^{-7}$ Torr). Cross-sectional TEM lamellas of the FET samples were prepared using a Helios 5 UX focused ion beam. The cross-sectional TEM images of a monolayer $MoS_2$ were taken at 200 keV using a FEI Talos F200X G2.

**Materials and process characterization**

Characterizations were carried out using scanning electron microscopy (Zeiss Sigma 550), optical microscopy (Olympus), AFM (Oxford, Jupiter XR), confocal microscope (Olympus, FV3000-BX63), Micro-focus X-Ray Photoelectron Spectroscopy (Thermo Fisher, Nexsa G2), confocal raman imaging microscopes (WITec, Alpha300RAS).

**Acknowledgments**

We thank Dr. Bowen Zhu and Dr. Xiaorui Zheng for discussions. We acknowledge the financial support from National Natural Science Foundation of China (61927820). We thank Jingjie Yan, Dr. Pengfei Wang, Abner-Nano Tech. LLC, and Shanghai OnWay Technology Co., Ltd for exfoliating the monolayer $MoS_2$. We thank the Instrumentation and Service Center for Physical Sciences at Westlake university for technical assistance. We acknowledge support from the Westlake Center for Micro/Nano Fabrication.

**Author contributions**

Min Qiu leads the whole research project. Xinyu Sun, BinBin Jin, Ding Zhao and Min Qiu conceived the main conceptual ideas. Ding Zhao and Min Qiu leads the fabrication of the self-designed system for ice-assisted process. Xinyu Sun leads the device fabrication and electrical characterization. BinBin Jin synthesized the perovskites. Wei Kong, Xiang Xu and Jichuang Shen synthetic the CVD-grown $MoS_2$. Wei Kong and Xiang Xu transferred and patterned CVD-grown $MoS_2$ onto target substrate. Yihan Lu, Jichuang Shen and Xinyu Sun contributed to the PL and XPS measurements. Yihan Lu, Xiang Xu and Xinyu Sun performed AFM measurements. Yihan Lu performed opto-electric measurements for the perovskites photodetector. Xinyu Sun analyzed the data. Xinyu Sun, Ding Zhao, and BinBin Jin co-wrote the manuscript. All authors contributed to the discussion of the manuscript.